\documentstyle[prl,aps,amsfonts,preprint,epsfig]{revtex}
\begin{document}
\draft
\tighten
\title{Correlated percolation and the correlated resistor network}
\author{Paul J. M. Bastiaansen\cite{paul} and Hubert J. F. Knops}
\address{Institute for Theoretical Physics,
University of Nijmegen, Toernooiveld, 6525 ED Nijmegen, The Netherlands}
%\date{}
\maketitle

\begin{abstract}
We present some exact results on percolation properties of the Ising
model, when the range of the percolating bonds is larger than
nearest-neighbors. We show that for a percolation range to next-nearest
neighbors the percolation threshold $T_p$ is still equal to the Ising
critical temperature $T_c$, and present the phase diagram for this type
of percolation. In addition, we present Monte Carlo calculations of the
finite size behavior of the correlated resistor network defined on the
Ising model. The thermal exponent $t$ of the conductivity that follows 
from it is found to be $t=0.2000\pm0.0007$. We observe no corrections to
scaling in its finite size behavior.
\end{abstract}


\section{Introduction}

The connection between percolation and the Ising model has been a
popular subject for a long time. One considers so-called Ising clusters
made up of nearest-neighbor spins with the same spin value. The
connectivity behavior of these clusters is called correlated site
percolation, as the probability distribution of the percolating and
non-percolating sites is a correlated one.

The interest in this problem arose because these Ising clusters were
believed to have the same properties as the droplets in the droplet
model\cite{fisher}, i.e.\ they should diverge at the Ising critical
point with the same critical exponents as those of the Ising model. It
became clear that they did indeed diverge\cite{coniglio77} at the Ising
critical point, but not\cite{sykes} with Ising exponents. An
alternative cluster definition was needed to have clusters with the
properties of droplets in the droplet model. These clusters, called
Coniglio-Klein clusters\cite{coniglio80}, are defined by putting bonds
between each pair of nearest neighbor up-spins, but now with a
probability $p=1-\exp(-2K)$, where $K$ is the Ising coupling. Not all
bonds of the Ising clusters appear in the Coniglio-Klein clusters, such
that the latter are, in that sense, `smaller' than Ising clusters. The
Coniglio-Klein clusters display\cite{coniglio80} the right critical
behavior: their diverge at the Ising critical point, their linear size
diverges as the Ising correlation length, and the mean cluster size
behaves as the susceptibility.

Both the Ising clusters and the Coniglio-Klein clusters have their
percolation point at the Ising critical temperature, albeit with
different critical behavior. The full picture of this cluster behavior
emerged\cite{murata,stella} when the behavior of both types of clusters
was identified with the phase diagram of the $q$-state dilute Potts
model in the limit $q\rightarrow1$. The tricritical point in this phase
diagram describes the behavior of the Ising clusters. This tricritical
point falls in the same universality class as the Ising critical
point\cite{nienhuis} in the sense that the central charge is $c=1/2$, but 
the critical exponents involved in the behavior of the Ising clusters
do not fit into the Ka\v{c}-table\cite{cardy}; they correspond to
half-integer values of the unitary grid. The Coniglio-Klein clusters
are described by the $1+1$ state symmetric fixed point in this phase
diagram.

Our motivation to reconsider the problem of correlated site percolation
is not the droplet theory of the Ising model but arises from a study of
correlated resistor networks; see further below. In this paper, we
consider Ising clusters that are made up of bonds with a larger
percolation range, that is, bonds are placed between nearest neighbor
up spins, but also between next- and further-neighboring pairs of equal
spins. Let us, throughout this paper, denote the clusters consisting of
bonds between nearest and next-nearest neighbor spin pairs by
nnn-clusters. In this language the Ising clusters are nn-clusters.

It is immediately clear that, if one considers bonds with a longer and 
longer percolation range, clusters get bigger and the percolation
threshold eventually will move to a temperature $T_p$ that is lower than
$T_c$, the Ising critical temperature. In that case, the type of
correlation is expected to be random percolation. In the limit of
percolating bonds with an infinite range, the percolation temperature
moves to $T_p=0$ and there is crossover to classical critical
behavior\cite{gouker}. In three dimensions this effect of a shift in
the percolation threshold occurs already with Ising clusters
(nn-clusters):  the percolation threshold lies at a temperature a few
percent below $T_c$\cite{muller}, whereas the Coniglio-Klein clusters
have their percolation threshold at $T_c$. In two dimensions it is known,
as stated above, that for nn-clusters $T_p$ coincides with $T_c$ but it
was believed\cite{coniglio82} that already for nnn-clusters $T_p<T_c$.
However, in this paper we shall show that $T_p=T_c$ for nnn-clusters.

These alternative cluster definitions can be useful in some
applications of correlated percolation. In another paper\cite{wij}, we
present a model, based on correlated percolation, to explain the
experimental results for colossal magnetoresistance. The latter
phenomenon is presently a hot topic in solid state physics\cite{cmr}.
Our model is a correlated resistor network, obtained by replacing
bonds with resistances yielding an effective resistance as an Ising
expectation value. In the present work, we present the technical
analysis of the correlated percolation model with percolating bonds
having a longer range of interaction. In particular, the resulting
phase diagram is vital for understanding the experimental results of
colossal magnetoresistance. The correlated resistor network has, to our
knowledge, never been studied in the literature. We performed Monte
Carlo calculations to measure the critical exponents of the CRN. Also
these calculations are presented in this paper.

\section{The model}

We will be concerned with the usual Ising model on a square lattice with 
Hamiltonian
\[
   H = -K\sum_{\langle ij\rangle} S_iS_j - h\sum_j S_j,
\]
where $K$ is the inverse temperature and $h$ is the magnetic field. The
first summation is over nearest neighbors only. Note that we are going
to define percolating bonds that have a longer range than nearest
neighbors only; the Ising Hamiltonian however, interacts only via
nearest neighbors.

\begin{figure}
   \begin{center}
   \epsfig{file=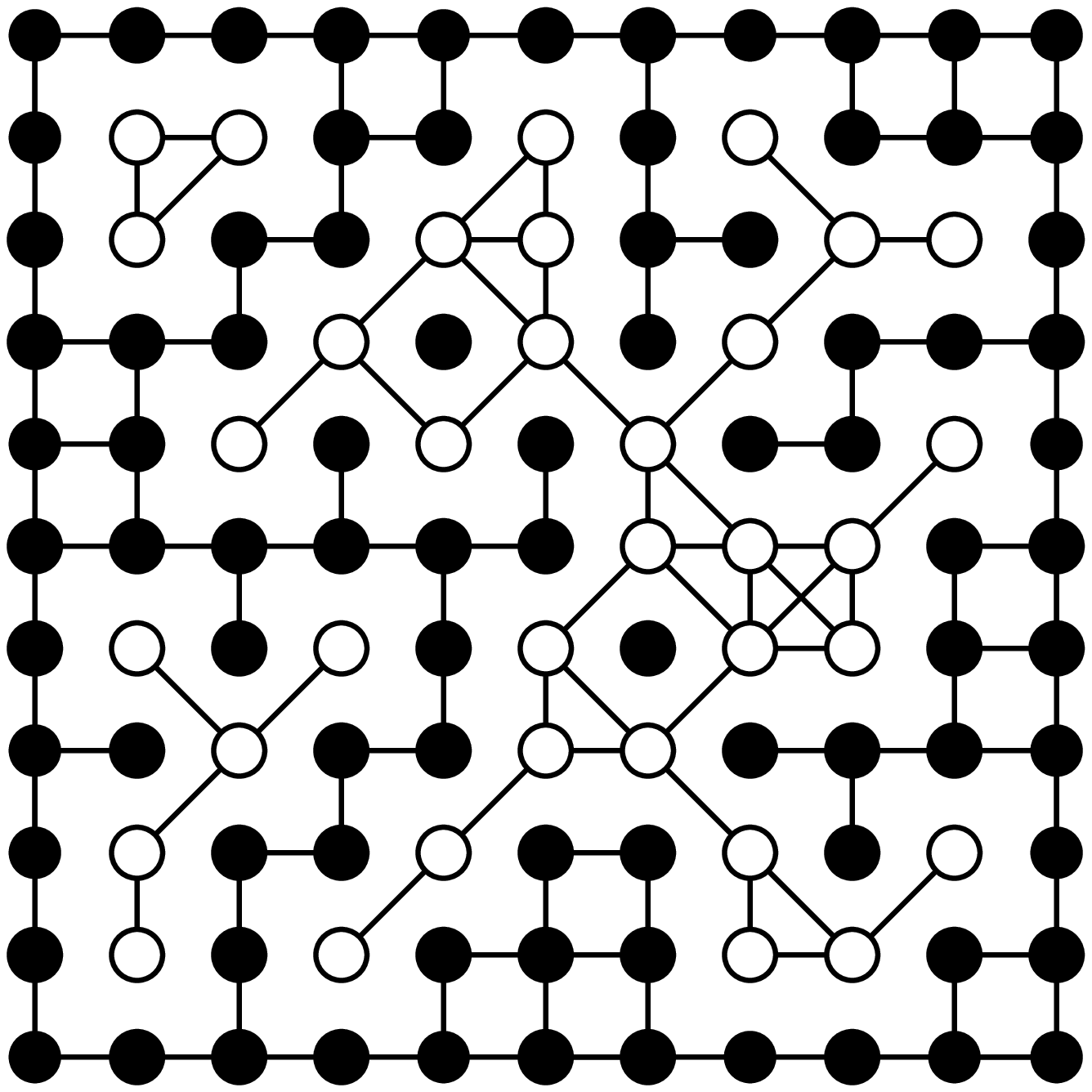,width=0.7\textwidth}
   \end{center}
   \caption{Illustration of the theorem presented in the text. On the
   sites of a lattice, black and white spins are placed. Spins on the
   boundary are always black. Bonds between black spins are present
   between each pair of black spins that are nearest neighbors. The
   same goes for the white spins, but here bonds are drawn if the spins
   are next-nearest neighbors as well. It can easily be inferred that a
   face of a black cluster is either empty or wholly occupied with one
   and only one white cluster. Notice that the inclusion of next-nearest
   neighbor bonds is essential.}
   \label{clusterconf}
\end{figure}

First consider the nn-clusters by putting bonds between all neighboring 
pairs of up spins, such as the clusters made of the black spins in
figure~\ref{clusterconf}. In the same figure, also the nnn-clusters are
illustrated but now for the white spins. Here bonds are put between
next-nearest neighboring pairs of white spins as well.

The percolation phase diagram for the nn-clusters is known\cite{stella}
and shown in figure~\ref{phasediag-nn}. The thick solid line is a
critical percolation transition; the end point at $T=\infty$
corresponds to random percolation and lies at the value of $h$
corresponding to the percolation threshold\cite{ziff} $p_c\approx0.5927$
for random percolation. This value denotes the density of up spins. The
corresponding value of $h$ is $h=0.188$. The critical percolation line
is in the universality class of random percolation, described by the
critical $q=1$ state Potts model. The line merges smoothly with the
$T$-axis at the Ising critical point. For percolation, this point
turns out\cite{stella} to be a tricritical point; it is the
tricritical point of the $q=1$ state dilute Potts model, where, apart
from the usual Potts spins, also vacancies are allowed.

From this phase diagram the corresponding diagram for nnn-clusters can
be derived. Figure~\ref{clusterconf} illustrates the theorem\cite{essam}
that will be needed for this derivation. It can easily be seen from
this figure that there exists a geometrical relation between the black
nn-clusters and the white nnn-clusters. The theorem states that {\it
every face of a black nn-cluster is either empty or is wholly occupied
by one and only one white nnn-cluster.} A face is a closed area
surrounded by elementary loops consisting of the bonds of one cluster.
Let us introduce the following notation: let $f_B$, $c_B$, $b_B$ and
$n_B$ be the number of faces, clusters, bonds and sites respectively
that correspond to the black spins and nn-clusters in a certain spin
configuration. For the white spins and nnn-clusters, these quantities
are $f_W^*$, $c_W^*$, $b_W^*$ and $n_W^*$ respectively, where the star
denotes the fact that it concerns nnn-clusters. The theorem then states
that
\[
   f_B = f_B^{(0)} + c_W^*, 
\]
where $f_B^{(0)}$ denotes the number of empty black faces. The number
of faces, bonds, sites and clusters is furthermore coupled via Euler's
equation
\[
   f_B = b_B - n_B + c_B + 1.
\]
This relation is easily derived by induction and holds for non-cyclic
boundary conditions, e.g., all spins on the boundary are black.
Combining these two relations yields
\[
   c_W^* = c_B + b_B - n_B - f_B^{(0)} + 1.
\]
Thus this relation expresses the number of white nnn-clusters $c_W^*$
in terms of the number of black nn-clusters $c_B$. Apart from the
numbers of clusters it involves only locally defined quantities: bonds,
sites and empty faces.

The above relations are purely geometrical and hence are completely
independent of the probability distribution of the black and white
spins. Therefore they are valid as Ising expectation values as well:
\begin{equation}
   \langle c_W^* \rangle = \langle c_B \rangle + \langle b_B \rangle - 
	  \langle n_B \rangle - \langle f_B^{(0)}\rangle + 1,
   \label{zwartwit}
\end{equation}
where the brackets denote the expectation values. Note that the numbers
of bonds, sites and empty black faces are simply local Ising operators,
that is
\[
   \langle b_B \rangle = \frac14\sum_{\langle ij\rangle}
   \big\langle(1+Si)(1+Sj)\big\rangle,
\]
which amounts to counting the number of black bonds. In the same way
\[
   \langle n_B \rangle = \frac12 \sum_j \left(1+\langle S_j\rangle\right)
\]
counts the number of black spins and
\[
   \langle f_B^{(0)}\rangle = \frac1{16}\sum_{\langle ijkl\rangle}
	\big\langle(1+S_i)(1+S_j)(1+S_k)(1+S_l)\big\rangle,
\]
counts the number of empty, black faces, where $\langle ijkl\rangle$ denotes 
a summation over each elementary plaquette of the square lattice.

The expectation value of the number of clusters plays the role of the
free energy in a percolation problem\cite{essam}. It becomes
non-analytic at a percolation transition. From
equation~(\ref{zwartwit}) we see that $c_W^*$ can only become critical
when $c_B$ is critical, that is, when the black spins are at their
percolation threshold, or when the Ising expectation values become
non-analytic, that is, at the Ising critical point and at the
coexistence line $T<T_c$, $h=0$. This immediately yields the phase
diagram for percolation of the black nnn-clusters: it is the mirror
image of that of the nn-clusters with $h$ replaced by $-h$. This phase
diagram is shown in figure~\ref{phasediag-nnn}.

It is somewhat surprising that extending the range of percolating bonds
to next-nearest neighbors does not alter the percolation threshold at the
$T$-axis. It was expected\cite{coniglio82} that for larger ranges than
nearest neighbor percolating bonds the percolation temperature $T_p$
would become less than $T_c$. Due to the above geometrical relations
this is not the case. For even larger percolation ranges, however,
there are no such relations and we expect the percolation temperature
$T_p$ to drop below the Ising critical temperature $T_c$. 

The topology of the phase diagram then changes. There is no tricritical
point anymore, and the critical percolation line will end somewhere at
the $T$-axis below $T_c$. Beyond this point, there still is a first
order transition for percolation. We expect the percolation point $T_p$
in that case to be a critical endpoint. Such a point is
expected\cite{berker} to have in addition to the critical exponents of
the universality class of the critical line also a first order exponent
$y=2$. Indeed, when the range of percolating bonds becomes very large,
it will eventually become larger than the Ising correlation length. In
such a case, correlations in the probability distribution of empty and
occupied sites are only present on a much smaller scale than the range
of percolating bonds.  This strongly suggests that the universality
class of the endpoint of the critical line is that of random
percolation.

Eventually, when the percolation range approaches infinity, percolation is 
believed to exhibit classical critical behavior\cite{gouker}.

\section{Scaling analysis}

From universality, we expect both types of critical behavior,
nn-percolation and nnn-percolation, to be in the same universality
class. In the light of equation~(\ref{zwartwit}), this statement is
less obvious than it seems. The singular behavior of the `free energy'
$c_W^*$ of nnn-clusters is expressed in $c_B$ but also in Ising
operators. Hence in addition to the critical behavior of $c_B$ also
Ising exponents show up. In particular, when $c_W^*$ is considered as a
function of the scaling field $u_1$, in addition to the exponent $2/y$
that is expected to describe the non-analytic behavior of the free
energy $c_B$, also the magnetization exponent $1/8$ is present.

Nn-percolation is in the universality class of the dilute $q$-state
Potts model in the limit of $q\rightarrow1$. Its
Hamiltonian\cite{murata,stella} is
\begin{equation}
   H = -L\sum_{\langle ij\rangle} n_in_j - \Delta\sum_j n_j -
	  J \sum_{\langle ij\rangle} n_in_j(\delta_{\sigma_i,\sigma_j}-1)
	  - H \sum_j n_j(\delta_{\sigma_j,1}-1).
   \label{Potts}
\end{equation}
The variables $n_i = 0,1$ are the Potts lattice gas variables. Potts
spins $\sigma_i=0,\ldots,q$ are present on the sites where $n_i=1$.

For $q=1$ the Hamiltonian becomes equal to the Ising Hamiltonian and
completely independent of $J$ and $H$. Substituting $S_i=2n_i-1$ turns
the lattice gas variables into Ising variables. The Hamiltonian then
becomes, apart from a constant,
\[
   H = -K\sum_{\langle ij\rangle}S_iS_j - h\sum_j S_j,
\]
with $K=L/4$ and $h=2L+\Delta/2$. This means that the free energy
resulting from equation~(\ref{Potts}) for $q=1$ is equal to that of the
Ising model. The full free energy is
\[
   f(L,\Delta,J,H,q) = -\lim_{N\rightarrow\infty}\frac1N
	  \ln Z^{(N)}(L,\Delta,J,H,q)
\]
where $N$ denotes the number of sites on the lattice. The generating
function\cite{fortuin} for percolation is
\[
   c(L,\Delta,J,H) = \left.
   \frac{d f(L,\Delta,J,H,q)}{dq} 
   \right|_{q=1}.
\]
This quantity $c$ is, with $H=0$, the expectation value of the number of 
clusters. It plays the role of the free energy in percolation problems. The
limit $J\rightarrow\infty$ corresponds to nn-clusters, and $J=2K$ yields
the Coniglio-Klein clusters.

In the language of the renormalization group, the tricritical point in
the $q=1$ dilute Potts model has four relevant exponents and
corresponding scaling fields, two thermal and two magnetic ones.
Expressing the free energy in terms of these scalings fields $u_i$, the
scaling relation is
\begin{equation}
   f(u_1,\ldots,u_4,q) = b^{-2} f(b^{y_1}u_1,\ldots,b^{y_4}u_4,q),
   \label{scaling}
\end{equation}
where $b$ is the renormalization length, and the tricritical point is
located at $u_1=\cdots=u_4=0$. Note that the `field' $q$, which is a 
symmetry parameter, cannot change under renormalization. Differentiating
this free energy with respect to $q$ yields the percolation free energy.
Applying this to the above scaling relation yields, apart from the
direct derivative to $q$, also derivatives with respect to the thermal
scaling fields $u_1$ and $u_2$ which are
\begin{eqnarray}
      u_1 =& K - K_c(q) + \cdots \\
      u_2 =& h - h_c(q) + \cdots
\end{eqnarray}
when $J=\infty$. The location of the tricritical point thus depends on
the value of $q$. The remaining two fields $u_3$ and $u_4$ are magnetic
scaling fields; they correspond to Potts-like magnetic fields. The
Potts-model can only be critical if those fields are zero, regardless
of the value of $q$, which means that the derivative of these fields
with respect to $q$ yields zero.

From these remarks it follows that in the expression for $c$ several
derivatives are present: a direct derivative with respect to $q$ and
derivatives with respect to the thermal scaling fields:
\begin{equation}
   c = \left.\frac{\partial f}{\partial q} +
  \frac{\partial f}{\partial u_1} \frac{\partial u_1}{\partial q} +
  \frac{\partial f}{\partial u_2} \frac{\partial u_2}{\partial q}
  \right|_{q=1}.
   \label{deriv}
\end{equation}
Applying equation~(\ref{scaling}) to this expression and taking the
limit of $q\rightarrow1$ yields the critical behavior of $c$. Note that
the two last terms of the right hand side of (\ref{deriv}) are with
$q=1$ just derivatives of the Ising free energy: they give the Ising
energy and magnetization, and thus yield Ising critical behavior. The
values of the exponents $y_1$ and $y_2$ are the Ising values $y_1=1$
and $y_2=15/8$. This gives for the singular behavior of $c$ as a
function of the thermal fields $u_1$
\begin{equation}
   c \sim A_1 |u_1|^{2/y_1} + A_2 |u_1|^{(2-y_1)/y_1}
	  + A_3 |u_1|^{(2-y_2)/y_1}.
   \label{scalingresult}
\end{equation}
The most relevant of these exponents is $(2-y_2)/y_1=1/8$, the exponent
of the Ising order parameter. The singular behavior of $c$ thus is
described by an exponent $1/8$, contrary to what is expected for the
behavior of a free energy-like quantity: the value of $y_1=1$ suggests a
critical exponent $2/y_1=1$ for the percolation free energy $c$. This
exponent indeed occurs in the expression for $c$ but is dominated by the
exponent $1/8$. This is due to the fact that $c$ is not a true free
energy but a derivative of a free energy with respect to a symmetry
parameter.

The order parameter $P$ (the density of sites in the infinite cluster)
and the `susceptibility' (the mean cluster size) arise from
differentiating $c$ with respect to $H$. The free energy $f$ itself becomes 
completely independent of $H$ in the limit for $q\rightarrow1$, which means 
that only the derivatives of $\partial f/\partial q$ with respect to
$H$ yield a non-zero result. The other terms in equation~(\ref{deriv})
vanish upon taking the derivative. This means that the critical
behavior of the order parameter and the susceptibility is not affected
by the `wrong' critical behavior of $c$.

From the scaling results (\ref{scalingresult}) of $c$ we can derive the
critical behavior of the nnn-clusters using equation~(\ref{zwartwit}).
The dominating exponent in the right hand side of (\ref{zwartwit}) is
again $1/8$, which appears in $c$ but also in the Ising magnetization
$\langle n_B\rangle$. This shows that the critical behavior of $c$ and
$c^*$, that is, of nn-clusters and nnn-clusters, is governed by the
same exponents.

The same must be true for the order parameter $P$ and the
susceptibility of nnn-percolation. Again these quantities arise from 
differentiating the free energy $c^*$ with respect to the Potts field $H$.
Although $c^*$ is different from $c$ in its dependence on $H$, both 
amount to a dilute $q=1$ states Potts model. With the identification of
the critical behavior of $c$ and $c^*$ it follows that at $T_c$ both
free energies must be at a tricritical point in the full phase diagram
of this model.

We conclude that percolation of nn-clusters and of nnn-clusters is in
the same universality class. The critical behavior of the percolation
quantities is governed by the same set of critical exponents.

\section{The correlated resistor network}

\begin{figure}
   \begin{center}
   \epsfig{file=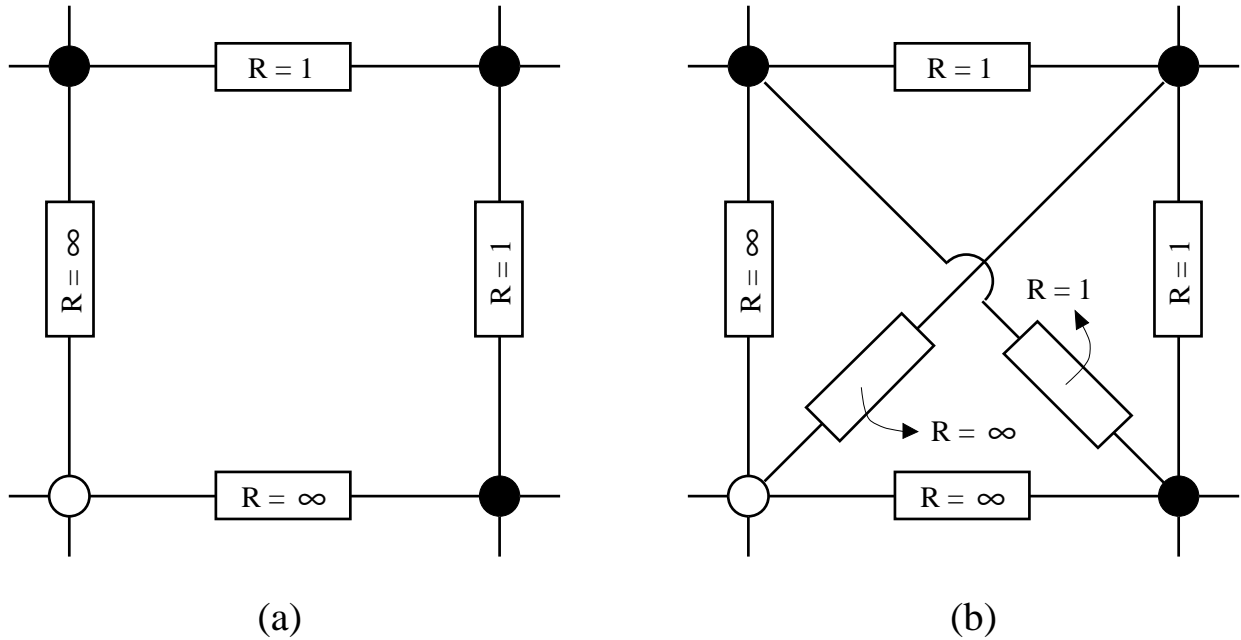,width=\textwidth}
   \end{center}
   \caption{The assignment of resistors to a spin configuration.
   Assignment for nn-clusters is shown in (a), for nnn-clusters in
   (b).  Here black spins (spin up) are considered percolating, so
   bonds between two black spins have a unit resistance. Bonds between
   black and white spins or between two white spins get an infinite
   resistance.}
   \label{resistors}
\end{figure}

Relevant for our work on colossal magnetoresistance\cite{wij} is the
behavior of the Ising model as a resistor network. A percolation model
is turned into a resistor network\cite{kirkpatrick} by replacing the
bonds with resistors. Non-percolating bonds are left empty (that is,
have infinite resistance), bonds that are present get a unit
resistance. This can be done both for nn-clusters and nnn-clusters.
The assignment of resistors is depicted in figure~\ref{resistors}.

For random percolation the resistor network is a random resistor network. 
The corresponding correlated resistor network has as far as we know
never been studied. In this section we present our calculations on the
correlated resistor network. The interest in resistor network problems
is in the expectation value of the overall resistance of the (infinite)
lattice. To be more precise:  consider a lattice consisting of $L\times
L$ spins, where $L$ eventually is sent to infinity, and keep the lower
row at a fixed potential $V=0$ and the upper row at $V=1$. The interest
is in the overall conductance $\sigma$ of the lattice, which is in this
case equal to the expectation value of the current.

\begin{figure}
   \begin{center}
   \epsfig{file=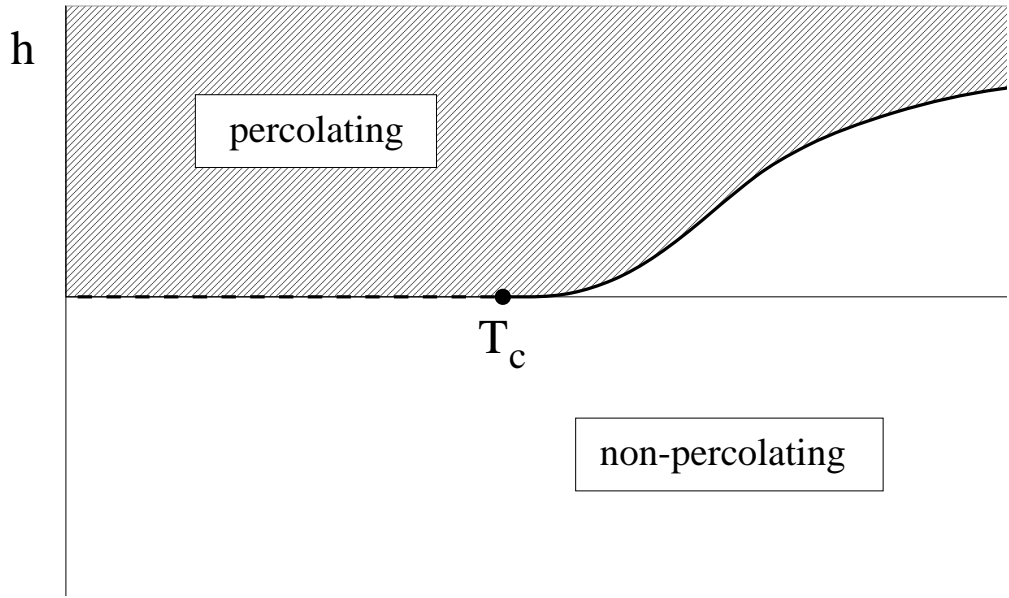,width=\textwidth}
   \end{center}
   \caption{The phase diagram for percolation of Ising clusters. Up
   spins are considered percolating, and clusters are defined by putting
   bonds between each neighboring pair of up-spins. $T_c$ is the Ising
   critical point, and $T$ and $h$ are temperature and magnetic field,
   respectively. The thick, solid line is a critical line of
   percolation that is in the universality class of random
   percolation.  $T_c$ is a tricritical point for percolation, and the
   dotted line is a first-order transition, for Ising as well as for
   percolation.}
   \label{phasediag-nn}
\end{figure}

\begin{figure}
   \begin{center}
   \epsfig{file=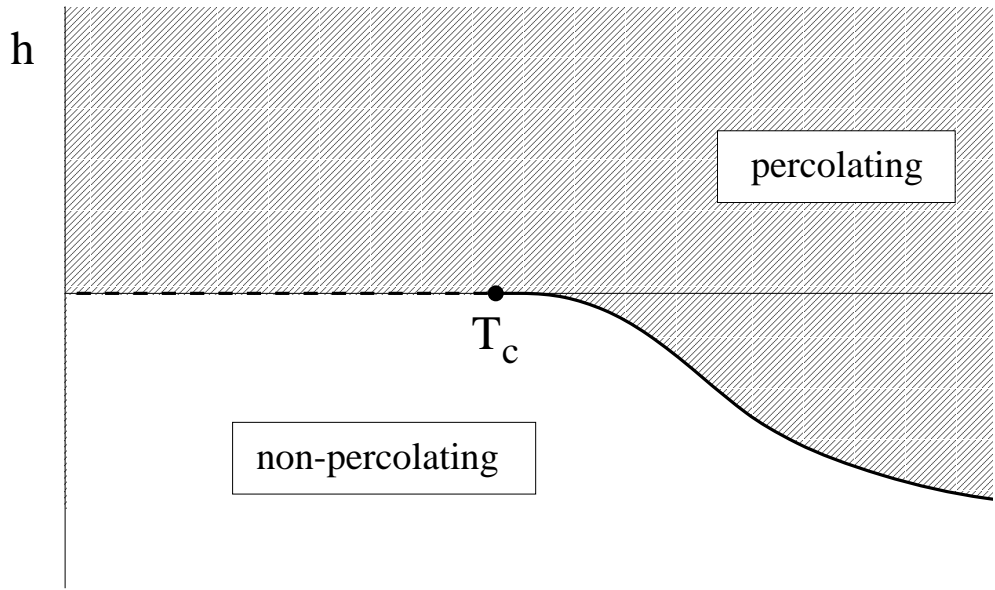,width=\textwidth}
   \end{center}
   \caption{The same phase diagram as in figure~\ref{phasediag-nn}, but
   now for nnn-clusters. Percolating bonds are drawn between pairs of up
   spins that are nearest or next-nearest neighbors. The phase diagram
   is the same as that for nn-clusters but with $h$ replaced with
   $-h$.}
   \label{phasediag-nnn}
\end{figure}

The phase diagram of the resistor network is of course the same as that
of its percolation counterpart: when there is percolation, the
conductance is finite, and the conductance is zero when there is no
percolation. Experimental results for colossal magnetoresistance show a
sharp increase in the resistance as a function of temperature at or
nearby the Curie point $T_c$. The resistance drops sharply both above
and below this point. From our phase diagrams it is directly clear that
the diagram of figure~\ref{phasediag-nn} is ruled out. So conduction via
next-nearest neighbor bonds should at least be present to produce the
right phase diagram.

Turning, however, to critical exponents the situation is different. In
this case it is the exponent $t$ governing the vanishing conductance
$\sigma$ upon approaching the percolation threshold:
\[
   \sigma(T) \sim |T-T_p|^t \quad\text{for $T\rightarrow T_p$}.
\]
Based on universality (again confirmed for the percolation exponents)
one may well assume that the exponent $t$ is the same for nn- and
nnn-networks. Hence we studied the (simpler) case of the resistor
network with only nn-clusters.

The random resistor network is a notoriously unsolved problem in
statistical physics, but there are good numerical estimates of the
exponent $t$. The best estimate\cite{normand} in two dimensions known
to us is $t=1.299\pm0.002$. To obtain the value of $t$ for the
correlated resistor network in the Ising case, at the tricritical
percolation point, we performed Monte Carlo calculations at the Ising
critical point, and calculated the Ising expectation value of the
conductance for different system sizes. We used the
Wolff-algorithm\cite{wolff} for the Monte Carlo part, and the multigrid
method of Edwards, Goodman and Sokal\cite{sokal}, based on the standard
code {\sc amg1r4}\cite{amg1r4}, to calculate the conductance of a spin
configuration.

\begin{figure}
   \begin{center}
   \epsfig{file=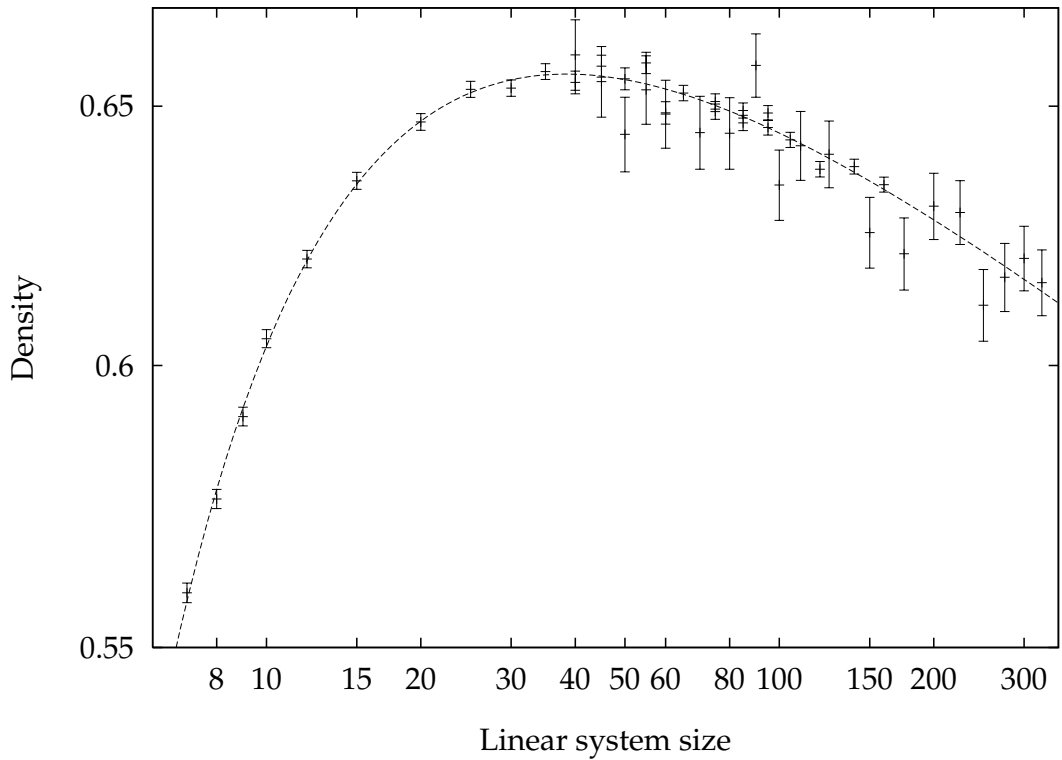,width=0.9\textwidth}
   \end{center}
   \caption{Log-log plot of the density of sites $P$ in the spanning
   cluster at the Ising critical point, as a function of the linear
   system size $L$.  The plot shows a far from straight line, meaning
   that corrections to scaling are important. The dashed line is the
   result of our fit against the function $P(L) =
   a_1L^{-\alpha}(1+a_2L^{-\beta})$. The values of $\alpha$ and $\beta$
   are $\alpha=0.055\pm0.005$ and $\beta=0.92\pm0.07$.}
   \label{density}
\end{figure}

To test our program, we calculated the exactly
known\cite{stella,nienhuis} exponent of the order parameter. The order
parameter of percolation is the density of sites $P$ in the infinite
cluster. From finite size scaling, it follows that this quantity scales
with the linear system size $L$ as
\begin{equation}
   P(L) \sim L^{-2+y_h} \quad\text{for $L\rightarrow\infty$},
   \label{fss-density}
\end{equation}
where $y_h$ is the most relevant magnetic eigenvalue. In our Monte
Carlo runs, we measured the number of sites in the `spanning cluster',
the cluster that extends over the lattice and thus allows for
conductance. With the scaling equation, the behavior of $P$ as a
function of the system size yields an estimate of $y_h$. We calculated
$P$ with the system size $L$ running from 7 up to 350, and the data are
plotted in figure~\ref{density}. The figure, a log-log plot, shows that
the system sizes are too small to exhibit the behavior of
equation~(\ref{fss-density}); corrections to scaling have to be
included.  We did this, and found $y_h=1.945\pm0.005$ and a correction
to scaling exponent $\omega=0.96\pm0.08$. The exact result is
$y_h=187/96\approx1.948$.  Our estimate thus agrees well within the
error bars.

\begin{figure}
   \begin{center}
   \epsfig{file=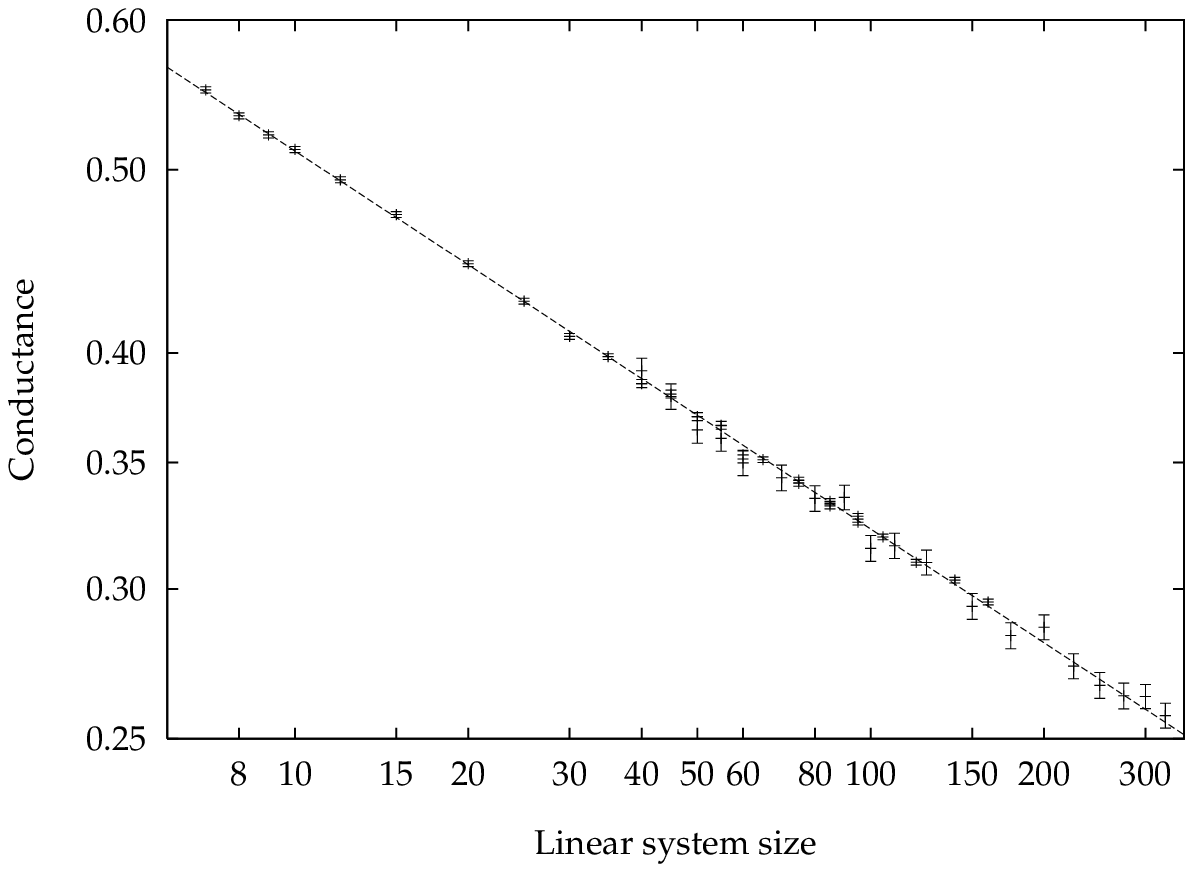,width=0.9\textwidth}
   \end{center}
   \caption{Log-log plot of the conductance $\sigma$ of the lattice at
   the Ising critical point, as a function of the linear system size
   $L$. The plot shows almost a straight line. Fitting the scaling
   behavior with a correction to scaling term did not yield sensible
   results. The dashed line is the result of our fit against the
   function $\sigma(L)=aL^{-\alpha}$, giving a value of
   $\alpha=0.2000\pm0.0007$.}
   \label{conductance}
\end{figure}

It is believed that a similar finite size scaling relation is valid for
the conductance $\sigma$. It should scale with the linear system size
$L$ as
\begin{equation}
   \sigma(L) \sim L^{-t/\nu},
   \label{fss}
\end{equation}
where $\nu$ is the percolation exponent of the correlation length. We
calculated the conductance again with $L$ running from 7 up to 350.
The data are plotted in figure~\ref{conductance}; the log-log plot almost
shows a straight line. In fitting the data to equation~(\ref{fss}), we
tried to include a correction to scaling term, but, due to the almost
perfect scaling behavior, this did not yield sensible results.
Therefore we performed the fit against equation~(\ref{fss}) without
additional terms, yielding the value $t/\nu = 0.2000\pm 0.0007$. Due to
the absence of the correction to scaling term the error in this result
might be an underestimation of the actual error.

The exponent $\nu$ of percolation at the Ising critical point is $\nu=1$
for the direction parallel to the $T$-axis and $\nu=8/15$ for the other 
directions. That means that the exponent $t$ that governs the vanishing 
conductance at $T_c$ is $t\approx 0.200$ for the temperature direction
and $t\approx 0.107$ for the field direction. This is a surprisingly low
result, as compared to the $t$ value of the random resistor network, 
$t\approx1.30$. The presence of critical correlations thus strongly 
influences the value of the conductance exponents.

Equation~(\ref{fss}) relies on the validity of finite size scaling, and
it is in this case not {\it a priori} clear that it is valid, since
there does not exist a rigorously defined renormalization
transformation for the conductance. Therefore we checked the
consistency of the value of the exponent by calculating the behavior of
the resistance in the neighborhood of $T_c$. The conductance then
scales as $\sigma \sim |T-T_c|^t$. The value of $t$ obtained from this
behavior was indeed seen to extrapolate to the value $0.20$. We believe
therefore that equation~(\ref{fss}) must be valid.

\begin{acknowledgments}
We thank Alan Sokal for providing us with the code for the resistance
calculations, which saved us a lot of work; Bernard Nienhuis for
discussions on correlated percolation; and Erik Luijten for discussions
on the Monte Carlo part.
\end{acknowledgments}

 \clearpage
 \clearpage
 \clearpage


\begin{references}
\bibitem[*]{paul} Electronic address: {\tt paulb@tvs.kun.nl}
\bibitem{fisher}
   Fisher M E 1967 {\it Physics} {\bf 3} 255
\bibitem{coniglio77}
   Coniglio A, Rosanna Nappi C, Peruggi F and Russo L 1977 {\it 
   J. Phys.} A {\bf 10} 205
\bibitem{sykes}
   Sykes M F and Gaunt D S 1976 {\it J. Phys.} A {\bf 9} 2131
\bibitem{coniglio80}
   Coniglio A and Klein W 1980 {\it J. Phys.} A {\bf 13} 2775
\bibitem{murata}
   Murata K K 1979 {\it J. Phys.} A {\bf 12} 81
\bibitem{stella}
   Stella A L and Vanderzande C 1989 {\it Phys. Rev. Lett.} {\bf 62} 1067
\bibitem{nienhuis}
   Nienhuis B 1987 in {\it Phase Transitions and Critical Phenomena} 
   vol 11, ed. C Domb and J L Lebowitz (New York: Academic) p 1
\bibitem{cardy}
   Cardy J L 1987 in {\it Phase Transitions and Critical Phenomena} 
   vol 11, ed. C Domb and J L Lebowitz (New York: Academic) p 55
\bibitem{gouker}
   Gouker M and Family F 1983 {\it Phys. Rev.} B {\bf 28} 1449
\bibitem{muller}
   M\"uller-Krumbhaar H 1974 {\it Phys. Lett.} {\bf 50A} 27
\bibitem{coniglio82}
   Jan N, Coniglio A and Stauffer D 1982 {\it J. Phys.} A {\bf 15} L699
\bibitem{wij}
   Bastiaansen P and Knops H (unpublished).
\bibitem{cmr}
   See, e.g., Kusters R M, Singleton J, Keen D A, McGreevy R and
   Hayes W 1989 {\it Physica} B {\bf 155} 362; 
   Chahara K, Ohno T, Kasai M and Kozono Y 1993 {\it Appl. Phys. 
   Lett.} {\bf 63} 
   1990; 
   McCormack M, Jin S, Tiefel T H, Fleming R M, Phillips J M and
   Ramesh R 1994 {\it Appl. Phys. Lett.} {\bf 64} 3045; 
   Schiffer P, Ramirez A P, Bao W and Cheong S-W 1995 {\it Phys. 
   Rev. Lett.} {\bf 75} 3336; 
   Shimakawa Y, Kubo Y and Manako T 1996 {\it Nature} {\bf 379} 53; 
   and references herein.
\bibitem{ziff}
   Ziff R M and Sapoval B 1986 {\it J. Phys.} A {\bf 19} L1169
\bibitem{essam}
   Essam J W 1972 in {\it Phase Transitions and Critical Phenomena} 
   vol 2, ed. C Domb and M S Green (New York: Academic) p 218
\bibitem{berker}
   Berker A N and Wortis M 1976 {\it Phys. Rev.} B {\bf 14} 4946
\bibitem{fortuin}
   Fortuin C M and Kasteleyn P W 1972 {\it Physica} {\bf 57} 536
\bibitem{kirkpatrick}
   See, e.g., Kirkpatrick S 1973 {\it Rev. Mod. Phys.} {\bf 45} 574
\bibitem{normand}
   Normand J-M, Herrmann H J and Hajjar M 1988 {\it J. Stat. Phys.} {\bf
   52} 441
\bibitem{wolff}
   Wolff U 1989 {\it Phys. Rev. Lett.} {\bf 62} 361
\bibitem{sokal}
   Edwards R G, Goodman J and Sokal A D 1988 {\it Phys. Rev. 
   Lett.} {\bf 61} 1333; 
   Goodman J and Sokal A D 1989 {\it Phys. Rev.} D {\bf 40} 2035
\bibitem{amg1r4}
   Ruge J and St\"uben K 1987 in {\it Multigrid Methods}, ed. S F McCormick
   (Philadelphia: SIAM)
\end{references}
\end{document}